\documentclass[aps,pra,superscriptaddress,twocolumn,10pt,nofootinbib,preprintnumbers,showdata]{revtex4-2}
\usepackage{bm}
\usepackage{epsfig}
\usepackage{graphics}
\usepackage{amsmath}
\usepackage{xcolor}

\begin{document}

\title{Weak decaying collective-excitation approximation \\ for Yukawa one-component plasmas}

\author{ \firstname{Ilnaz~I.}~\surname{Fairushin}}
\author{ \firstname{Anatolii~V.}~\surname{Mokshin}}

\affiliation{Department of Computational Physics, Institute of Physics, Kazan Federal University, 420008 Kazan, Russia}

\begin{abstract}
In this paper, the theoretical model of \textit{weak decaying} collective excitations characteristic of many-particle systems with long-range interaction potentials is developed using the example of one-component strongly coupled Yukawa plasmas. The proposed model is based on the self-consistent relaxation theory of collective dynamics and covers spatial scales from extended hydrodynamics to scales related to the mean interparticle distance. The theoretical model reproduces the dynamic structure factor spectra and the corresponding dispersion characteristics in agreement with molecular dynamics simulation data without using any fitting parameters. In the limit of small wave numbers, the correspondence of the proposed theoretical model with the damped harmonic oscillator model is established. The simple analytical expression for the sound attenuation coefficient of strongly coupled Yukawa plasmas is obtained.
\end{abstract}

\maketitle

\section{Introduction}
\subsection{Collective particle dynamics in simple liquids}
\begin{figure}[h!]
\includegraphics[width=8.0cm]{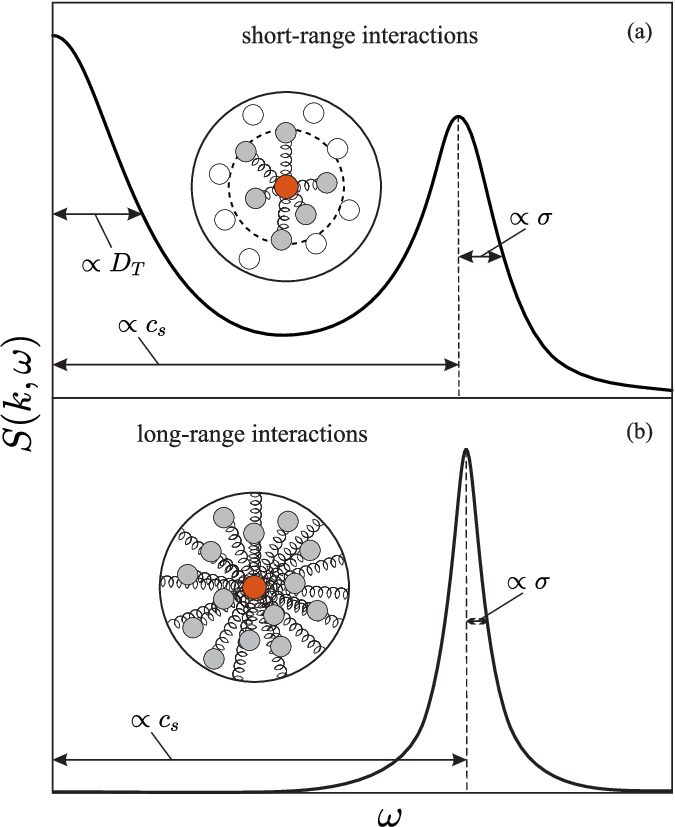}
\caption{Typical dynamic structure factor spectrum for an equilibrium simple liquid with short-range (a) and long-range (b) interaction potential. Here, $D_T$ is the thermal diffusivity, and $c_s$ and $\sigma$ are the adiabatic sound velocity and sound attenuation coefficient, respectively.} 
\label{Spectra_scheme}
\end{figure}

The collective particle dynamics in simple equilibrium liquids is characterized by translational and vibrational modes. At hydrodynamic scales, the translational mode provides heat transfer and the vibrational mode provides propagation of sound waves \cite{Balucani, Hansen/McDonald_book_2006, Boon/Yip_1991, Khrapak_review_2024, Zhou, Zaccone, Usov, Trigger}. These features in the collective particle dynamics of simple equilibrium liquids are manifested in its dynamic structure factor (DSF) spectra $S(k, \omega)$, where $k$ is the wave number and $\omega$ is the frequency. At small $k$, the DSF spectrum of an equilibrium simple liquid is characterized by central (Rayleigh) and side (high-frequency or Brillouin) peaks (Fig. \ref{Spectra_scheme}a). The Rayleigh peak is associated with the translational mode of collective particle dynamics, while the Brillouin peaks are associated with the vibrational mode of collective particle dynamics. The half width at half height of the central peak is proportional to the value of the thermal diffusivity $D_T$, and the half width at half height of the high-frequency peak is proportional to the value of the sound attenuation coefficient $\sigma$. The position of the high-frequency peak is proportional to the adiabatic sound velocity $c_s$ \cite{Balucani, Hansen/McDonald_book_2006, Boon/Yip_1991} (Fig. \ref{Spectra_scheme}a). 
The key factor that determines the collective particle dynamics, as a whole, and is responsible for the ratio of intensities between its translational and vibrational components, is a character of the interparticle interaction. The pronounced long-ranged interaction leads to the fact that local density fluctuations propagate rapidly throughout the system without appreciable transfer of thermal energy. In the $S(k,\omega)$ spectra, this is manifested in the fact that at small wave numbers the Rayleigh peak is practically absent, and the Brillouin peak is characterized by a small width (Fig. \ref{Spectra_scheme}b). Collective excitations with such features can, in fact, be labeled as \textit{weak decaying}. The many-particle systems with these features in the collective dynamics include the Yukawa one-component plasmas (YOCP) in the weak and intermediate screening regimes \cite{Barrat, Ichimaru, Hamaguchi, Fortov_reviews_1, Fortov_reviews_2, Killian_reviews, Kremer, Bonitz_PRL_1, Bonitz_PRL_2, Murillo, Fluids, Mithen_PRE_2011, Khrapak_entropy, KhrapakJCP2019, Baggioli, Arkhipov, TkachenkoPRE2020, Tkachenko_PRB_2023, BraultPRE2025, MurilloPRR2022, FM_PRE, MFT_PRE}. This behavior of YOCP is partly similar to the behavior of crystalline substances, where the translational component in the collective dynamics is completely absent and the vibrational component is characterized by a narrow high-frequency peak in the DSF spectra. In this connection, the so-called quasicrystalline approximation, known also as the quasilocalized charge approximation (QLCA), has been widely used in describing the collective dynamics of the YOCP particles \cite{Kalman, Golden, Khrapaks_Phys_Plasma_2015, Klumov, Fairushin}. 
\subsection{Features of collective excitations in the YOCP}
\begin{figure}[h!]
\includegraphics[width=8.65cm]{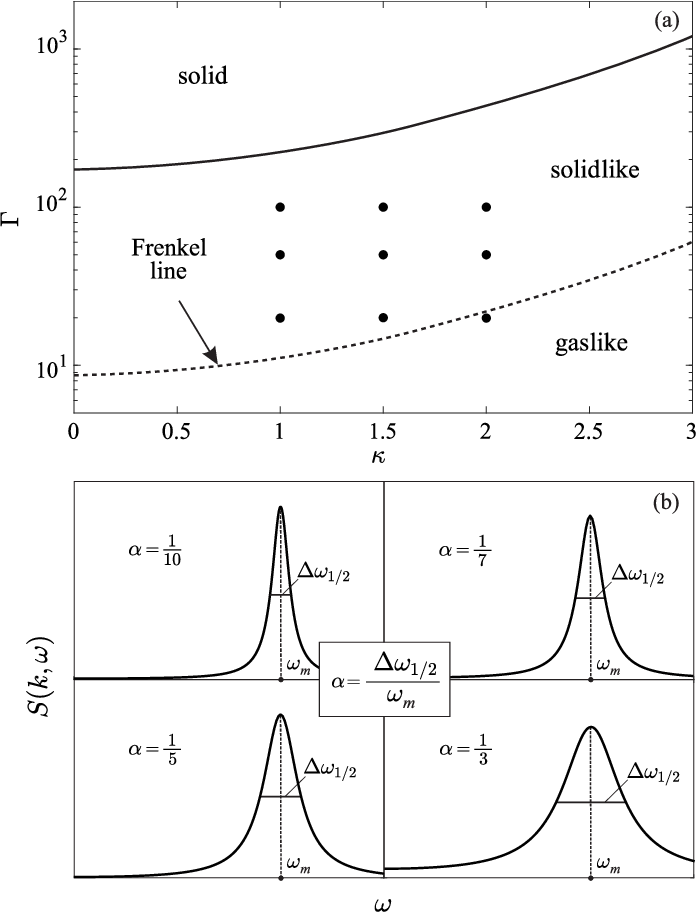}
\caption{a -- Phase diagram of the YOCP \cite{Hamaguchi, Baggioli} showing those ($\Gamma, \kappa$) states that are considered in this paper. b -- Typical DSF spectra for the YOCP in the regime of \textit{weak decaying} collective excitations at different values of the parameter $\alpha$, which is the ratio of the FWHM of the high-frequency peak $\Delta\omega_{1/2}$ to its position $\omega_m$.} 
%The inset illustrates the phase diagram of the YOCP \cite{Hamaguchi, Baggioli}, featuring the ($\Gamma, \kappa$)-states considered in this work.}
\label{Spectra_PD}
\end{figure}

In the present work, we develop the theoretical model for the DSF that describes \textit{weak decaying} collective excitations in the YOCP. The YOCP is a system of identical interacting particles with some mass $m$ and charge $Q$ immersed in a neutralizing background \cite{Barrat, Ichimaru, Hamaguchi, Fortov_reviews_1, Fortov_reviews_2, Killian_reviews, Kremer, Bonitz_PRL_1, Bonitz_PRL_2, Murillo, Fluids, Mithen_PRE_2011, Khrapak_entropy, KhrapakJCP2019,  Baggioli, Arkhipov, TkachenkoPRE2020, Tkachenko_PRB_2023, BraultPRE2025, MurilloPRR2022}. 
The YOCP is an actively researched model of strongly coupled plasmas \cite{Barrat, Ichimaru, Hamaguchi, Fortov_reviews_1, Fortov_reviews_2}. Part of the scientific interest in this model is due to the fact that this model can be used as a first approximation in describing the physical properties of such objects as dusty plasma \cite{Fortov_reviews_1, Fortov_reviews_2}, ultracold plasma \cite{Killian_reviews}, plasma in inertial confinement fusion devices \cite{Fortov_reviews_1}, and others. In contrast to the Coulomb one-component plasmas \cite{FM_PRE}, the thermodynamic state of the YOCP is given by two key parameters. First, there is the Coulomb coupling parameter,
\begin{equation}
\Gamma = \frac{Q^2}{4\pi\varepsilon_0ak_BT}\,,
\label{Gamma}
\end{equation}
which characterizes how many times the Coulomb interaction energy of particles at a distance equal to the radius of the Wigner-Seitz cell $a = (3/4\pi \rho)^{1/3}$ is greater than their thermal motion energy $k_BT$. Here, $\rho$ is the number density of the particles, $\varepsilon_0$ is the electric constant, $k_B$ is the Boltzmann constant and $T$ is the absolute temperature of the system. Second, there is the screening parameter,
\begin{equation}
\kappa=\frac{a}{\lambda_s},
\end{equation}
where $\lambda_s$ is the Debye screening length, which determines the softness of the Yukawa interaction potential:
\begin{equation}\label{Yukawa_pot}
u(r) = \frac{Q^2}{4\pi\varepsilon_0 r}\exp\left(-\frac{r}{\lambda_s}\right).
\end{equation}
At $\lambda_s\rightarrow\infty$, the Yukawa potential reduces to the known Coulomb interaction potential, and, at $\lambda_s\rightarrow0$, it approximates the hard-sphere interaction potential. Strong nonideality of the considered many-particle system corresponds to the coupling parameter $\Gamma>1$ \cite{Fortov_reviews_1, Fortov_reviews_2} (Fig. \ref{Spectra_PD}a). 
\textit{Weak decaying} collective excitations in the YOCP are observed over the range of wave numbers $0<k\lesssim k_m/2$, where $k_m$ is the wave number corresponding to the first maximum in the static structure factor of the system $S(k)$ \cite{MFT_PRE}. The $k$ dependence of high-frequency peak position $\omega_m(k)$ gives the dispersion of the longitudinal collective excitations. At small wave numbers, this dependence is linear, the slope of which determines the adiabatic sound velocity~$c_s$:
\begin{equation}\label{w_c_low_k}
\lim_{k\rightarrow0}\omega_m(k)=c_sk.
\end{equation}
The full width at half maximum (FWHM) $\Delta\omega_{1/2}$ of the high-frequency peak in the DSF spectrum (Fig. \ref{Spectra_PD}b) is related with attenuation decrement $\delta$ of longitudinal collective excitations: $\delta=-\Delta\omega_{1/2}/2$ \cite{Balucani, Hansen/McDonald_book_2006, Boon/Yip_1991}. The attenuation decrement depends on the properties of the medium, such as viscosity, thermal diffusivity, and the characteristic frequency $\omega_m$ of the collective excitations. It is known that the frequency dependence of the attenuation decrement $\delta$ has a quadratic law: $\delta \propto \omega_m^2$ \cite{Balucani, Hansen/McDonald_book_2006, Boon/Yip_1991}. Further, since at small $k$ the frequency of oscillations has dependence \eqref{w_c_low_k}, for the dispersion of the attenuation decrement of longitudinal collective excitations in this limit we obtain \cite{Balucani, Hansen/McDonald_book_2006, Boon/Yip_1991}
\begin{equation}\label{delta_low_k}
\lim_{k\rightarrow0}\delta(k)=-\sigma k^2.
\end{equation}
Comparing expressions \eqref{w_c_low_k} and \eqref{delta_low_k}, we see that the ratio of the FWHM $\Delta\omega_{1/2}$ to its position $\omega_m(k)$,
\begin{equation}\label{alpha}
\alpha = \frac{\Delta\omega_{1/2}}{\omega_m},
\end{equation}
is directly proportional to $k$. In the range of wave numbers $0<k\lesssim k_m/2$ for the YOCP with $\Gamma\in[20; 100]$ and $\kappa\in[1; 2]$, the parameter $\alpha$ does not exceed $1/3$ \cite{MFT_PRE}. Further, we will consider the states with the coupling and screening parameters from the specified ranges (Fig. \ref{Spectra_PD}a).

Since the ratio of intensities of the high-frequency $I_B$ and central $I_R$ peaks in the DSF at small wave numbers is related via the Landau-Plachek formula to the ratio $\gamma$ of specific heat capacities at constant pressure and volume \cite{Landau_Placzek},
\begin{equation}
\frac{I_R}{2I_B}=\gamma-1,
\end{equation}
for the many-particle systems in \textit{weak decaying} collective excitations regimes we have
\begin{equation}\label{gamma}
 \gamma=\frac{c_p}{c_v}\approx1,
\end{equation}
where $c_p$ and $c_v$ are the specific heat capacities at constant pressure and volume, respectively. When condition \eqref{gamma} is satisfied, taking into account relations \eqref{w_c_low_k} and \eqref{delta_low_k}, the following expression is obtained for the DSF of the YOCP in the hydrodynamic limit \cite{Mithen_PRE_2011}:
\begin{equation}\label{Skw_H}
S^\textrm{H}(k,\omega) = \frac{S(k)}{\pi}\frac{2(c_sk)^2\sigma k^2}{[\omega^2 - (c_sk)^2]^2+[2\omega\sigma k^2]^2}.
\end{equation}
Thus, this relation reproduces the DSF spectra of the YOCP at the low-$k$ limit, which contains a single high-frequency peak at frequency $\omega_m=c_sk$ with the FWHM $\Delta\omega_{1/2}=2\sigma k^2$ (Fig. \ref{Spectra_PD}b). At the same time, the following problem remains open: obtaining simplified analytical expressions for the characteristics of collective dynamics of a many-particle system with the features associated with the \textit{weak decaying} type of the acousticlike excitations. The present work is devoted to solving this problem.
% Note that when condition \eqref{gamma} is satisfied, the adiabatic sound velocity $c_s$ in the YOCP becomes equal to the isothermal sound velocity \cite{MokshinTMF}.

The paper is organized as follows. In Sec. II, we describe the theoretical formalism related with the self-consistent relaxation theory of collective dynamics and \textit{weak decaying} collective excitations approximation for the YOCP. In Sec. III, the obtained theoretical results are compared with molecular dynamics (MD) simulations data and the results of other theoretical approaches. The main findings are given in the Conclusion (Sec. IV).

\section{Theoretical formalism}
\subsection{Self-consistent relaxation theory}
According to the general definition, the DSF $S(k,\omega)$ is the temporal Fourier transform of the intermediate scattering function $F(k,t)$ \cite{Balucani, Hansen/McDonald_book_2006, Boon/Yip_1991}:
\begin{equation}
S(k,\omega) = \frac{S(k)}{2\pi}\int_{-\infty}^\infty F(k,t)\exp(\textbf{i}\omega t)dt\,,
\label{Skw_from_Fkt}
\end{equation}
where $t$ is the time.
Let us express the function $F(k,t)$ from Eq. \eqref{Skw_from_Fkt} and decompose it by time $t$ into the McLorean series:
\begin{equation}
\begin{gathered}
F(k,t) = 1 - \langle\omega^{(2)}(k)\rangle\frac{t^2}{2!}+\langle\omega^{(4)}(k)\rangle\frac{t^4}{4!}+\dots \\
+(-\textbf{i})^l\langle\omega^{(l)}(k)\rangle\frac{t^l}{l!}+\dots \,,
\label{Fkt_series}
\end{gathered}
\end{equation}
where $\langle\omega^{(l)}(k)\rangle$ is the normalized frequency moment of $S(k, \omega)$ of $l$th order ($l = 2, 4 , 6, ...$):
\begin{equation}
\langle\omega^{(l)}(k)\rangle = (-\textbf{i})^l\frac{d^l}{dt^l}F(k,t)\bigg|_{t=0}= \frac{\int_{-\infty}^\infty \omega^l S(k,\omega)d\omega}{{S(k)}}\,.
\label{nmoments_basic}
\end{equation}
From Eq. \eqref{Fkt_series}, one obtains the following expression for the Laplace transform of the function $F(k,t)$:
\begin{equation}
\begin{gathered}
\widetilde{F}(k, s) = \frac{1}{s} - \frac{\langle\omega^{(2)}(k)\rangle}{s^3} + \frac{\langle\omega^{(4)}(k)\rangle}{s^5} +\dots \\
+ (-\textbf{i})^l\frac{\langle\omega^{(l)}(k)\rangle}{s^{l+1}} +\dots\,,
\label{Fks_series}
\end{gathered}
\end{equation}
which can be rewritten as the following continued fraction:
\begin{equation}
\widetilde{F}(k,s) = \cfrac{1}{s+\cfrac{\Delta_1(k)}{s+\cfrac{\Delta_2(k)}{s+\cfrac{\Delta_3(k)}{s+\ddots}}}}
\,.
\label{Fks_cfrac}
\end{equation}
Here, $\Delta_n(k)$ are the so-called frequency relaxation parameters \cite{FM_PRE, Mokshin, AVM_PRE_2001, AVM_JCP, AVM_JPCM, Mokshin_PRB_2020, MokshinTMF, MFT_PRE}. Comparing Eqs. \eqref{Fks_series} and \eqref{Fks_cfrac}, we obtain the following relations between the frequency relaxation parameters and the frequency moments of $S(k,\omega)$:
\begin{eqnarray}
\Delta _{1}(k) &=&\frac{\langle \omega ^{(2)}(k)\rangle }{\langle \omega
^{(0)}(k)\rangle }, \\
\Delta _{2}(k) &=&\frac{\langle \omega ^{(4)}(k)\rangle }{\langle \omega
^{(2)}(k)\rangle }-\frac{\langle \omega ^{(2)}(k)\rangle }{\langle \omega
^{(0)}(k)\rangle },  \notag \\
\Delta _{3}(k) &=&\frac{\left[ \langle \omega ^{(6)}(k)\rangle \langle
\omega ^{(2)}(k)\rangle -\left( \langle \omega ^{(4)}(k)\rangle \right) ^{2}%
\right] \langle \omega ^{(0)}(k)\rangle }{\langle \omega ^{(4)}(k)\rangle
\langle \omega ^{(2)}(k)\rangle \langle \omega ^{(0)}(k)\rangle -\left(
\langle \omega ^{(2)}(k)\rangle \right) ^{3}},  \notag \\
\,\,\, \dots, \notag \\ 
\Delta _{n}(k) &=& \mathcal{F}\left[ \langle \omega ^{(0)}(k)\rangle, \langle
\omega ^{(2)}(k)\rangle, \dots , \langle \omega ^{(2n)}(k)\rangle \right],  \notag
\end{eqnarray}%
where $\mathcal{F}[\ldots]$ means an algebraic expression. Note that each frequency relaxation parameter is related to the corresponding dynamical variable from an orthogonal basis, the first element of which is the density fluctuations \cite{Mokshin, AVM_JPCM, AVM_PRE_2001, AVM_JCP, Mokshin_PRB_2020, MokshinTMF, MFT_PRE, FM_PRE}. In addition, the $n$-order frequency relaxation parameter is related to the corresponding distribution function of $n$ particles of the system and characterizes the oscillatory process for different groups of $n$ neighboring particles \cite{Mokshin, AVM_JPCM, AVM_PRE_2001, AVM_JCP, Mokshin_PRB_2020, MokshinTMF, MFT_PRE, FM_PRE}.

Various theoretical models for the DSF, including models for the YOCP, are based on different ways of truncating continued fraction \eqref{Fks_cfrac}. For example, the theory based on the frequency moments method allows one to calculate directly the DSF of the YOCP without any fitting parameters \cite{Arkhipov, TkachenkoPRE2020, Tkachenko_PRB_2023}. In this case, the expression for $S(k, \omega)$ is obtained as a result of a fractional-linear transformation using the Nevanlinna parameter function, which is one of the forms of representation of continued fraction \eqref{Fks_cfrac}. There are also such theoretical models for the DSF that are based on exponential \cite{Boon/Yip_1991} and Gaussian \cite{Mithen_PRE_2011} memory functions. However, these models contain different fitting parameters. In the present work, we develop the theoretical model for the $S(k, \omega)$ of the YOCP based on the self-consistent relaxation theory of the collective dynamics of many-particle systems \cite{Mokshin, AVM_JPCM, AVM_PRE_2001, AVM_JCP, Mokshin_PRB_2020, MokshinTMF, MFT_PRE, FM_PRE}. The key idea of self-consistent relaxation theory is as follows: if there is a certain correspondence between the time scales of the dynamic variables, the analytical solutions can be found for relaxation characteristics. The existence of such correspondence allows one to calculate a number of physical characteristics, such as spectral densities of time correlation functions of various physical quantities \cite{Mokshin, Mokshin_PRB_2020, Sawada_PRE_2010, Sawada_PRL_1999, Grebenkov_PRE_2014, Belyi_PRE_2018, Tong_PRB_2021}. As applied to simple liquids, the self-consistent relaxation theory is based on the fact that the time scales of dynamical variables above the fourth order are outside the processes associated with structure relaxation. In this case, the characteristic frequencies of dynamical variables beginning from the fourth order are aligned:
\begin{equation}
\label{Deltas_equal}
\Delta_4(k) = \Delta_5(k) = \Delta_6(k) = ... .
\end{equation}
These relations mean that it is necessary to know only the first four frequency relaxation parameters to calculate the $S(k, \omega)$ of simple one-component systems. The following analytical expressions for $\Delta_1(k)$ and $\Delta_2(k)$ are known for the YOCP \cite{Mithen_PRE_2011, MFT_PRE}:
\begin{equation}
\Delta_1(k) = \frac{\omega_p^2(ka)^2}{3\Gamma S(k)} \,,
\label{delta_1_micro}
\end{equation}
\begin{equation}
\Delta _{2}(k)=\Delta _{1}(k)\left[ 3S(k)-1\right] +\omega _{p}^{2}\,D(k),
\label{eq: Delta_2}
\end{equation}
where
\begin{equation}
\begin{gathered}
D(k) = \int_{0}^{\infty }\frac{\exp (-\kappa x)}{3x}\bigl[\left( 2(\kappa
x)^{2}+6\kappa x+6\right) j_{2}(kax) \\ +(\kappa x)^{2}(1-j_{0}(kax))\bigr]g(x)dx.  \notag
\end{gathered}
\end{equation}
Here, $\omega_p=\sqrt{Q^2\rho/(\varepsilon_0m)}$ is the plasma
frequency, $x=r/a$ is the dimensionless distance, $j_{n}(x)$ is the spherical Bessel functions of the first kind, and $g(x)$ is the radial distribution function. In addition, as was shown in Ref. \cite{MFT_PRE}, in the case of the YOCP, the third- and fourth-order frequency relaxation parameters can be expressed through the second-order parameter using the following correlation relations:
\begin{subequations}
\label{eq: freq_parameters_approx}
\begin{equation}
\Delta_3(k) \approx \frac{3}{2}\Delta_2(k) + \omega_0^2, 
\label{approx1}
\end{equation}
\begin{equation}
\Delta_4(k) \approx \frac{4}{3}\Delta_3(k) \approx 2\Delta_2(k) + \frac{4}{3}\omega_0^2,
\label{approx2}
\end{equation}
\end{subequations}
where
\begin{equation}
\omega_0^2 = \frac{2\,\omega_p^2}{\sqrt{\Gamma\kappa}}.
\nonumber
\end{equation}
Taking into account Eqs. \eqref{Deltas_equal}, \eqref{approx1}, and \eqref{approx2}, the following analytical expression for the $S(k, \omega)$ of the YOCP is obtained:
\begin{equation}\label{Skw_MSC}
S(k,\omega )=\frac{S(k)}{\pi }\frac{2\Delta _{2}(k)\sqrt{A_{3}(k)}}{\omega
^{6}+A_{1}(k)\omega ^{4}+A_{2}(k)\omega ^{2}+A_{3}(k)},
\end{equation}
where 
\begin{eqnarray}
A_{1}(k) &=&3\omega _{0}^{2}-\frac{\Delta _{2}(k)}{2}-2\Delta _{1}(k), 
\notag \\
A_{2}(k) &=&\left[ \Delta _{1}(k)-2\Delta _{2}(k)\right] ^{2}-6\Delta
_{1}(k)\omega _{0}^{2},  \notag \\
A_{3}(k) &=&\frac{3}{2}\Delta _{1}^{2}(k)\left( 3\Delta _{2}(k)+2\omega
_{0}^{2}\right) .  \notag
\end{eqnarray}
This equation is general for the liquid phase of the YOCP and reproduces the collective dynamics of ions over a wide range of wave numbers, where both the Rayleigh and Brillouin modes could be pronounced. At the same time, as was said above, in the YOCP, due to the long-range interaction of ions, longitudinal collective excitations with wave numbers in the range $0<k\lesssim k_m/2$ represent only propagating density fluctuations that gradually decay. This character of collective excitations corresponds to the damped harmonic oscillator (DHO) model. Then, the next reasonable questions arise. Can general solution \eqref{Skw_MSC} be transformed into a DHO-like model? How, in this case, will the characteristic parameters of the DHO model be determined: the characteristic frequency of propagating oscillations and the rate of their decay? On the other hand, how does the solution obtained within the self-consistent relaxation theory relate to the known solution \eqref{Skw_H} for the DSF spectra of the YOCP in the hydrodynamic regime?

\subsection{Theoretical model for the \textit{weak decaying} collective excitations in the YOCP}

We start our discussion with the well-known result of the QLCA model \cite{Kalman, Golden, Khrapaks_Phys_Plasma_2015, Klumov, Fairushin}. In the range of wave numbers $0<k\lesssim k_m/2$ the QLCA model with good accuracy reproduces the position of side peak in the DSF spectra of the YOCP through the following relation:
\begin{equation}\label{QLCA}
\omega_c^{\textrm{QLCA}}(k) = \sqrt{\Delta_1(k)+\Delta_2(k)}.
\end{equation}
This fact can be used to simplify expression (\ref{Skw_MSC}). The simplification is based on the fact that the DSF spectrum containing only the high-frequency peak can be described by a function that is a fourth degree polynomial (in frequency) (see Eq. \eqref{Skw_H}). Let us represent expression (\ref{Skw_MSC}) as some function of the square of the frequency, $\Omega=\omega^2$, i.e.,
\begin{equation}\label{f(x)}
S(k,\Omega) = \frac{A_4(k)}{\Omega^3+A_1(k)\Omega^2+A_2(k)\Omega+A_3(k)}.
\end{equation}
Here, $A_4(k)=2S(k)\Delta_2(k)\sqrt{A_3(k)}/\pi$. From Eq. \eqref{f(x)} we see that in the function $S(k,\Omega)$ only the denominator depends on $\Omega$. This denominator we will denote as
\begin{equation}
P(\Omega,k)=\Omega^3+A_1(k)\Omega^2+A_2(k)\Omega+A_3(k).
\end{equation}
The extremum condition on the variable $\Omega$ for both the function $S(k,\Omega)$ and the function $P(\Omega,k)$ is given by the following expression:
\begin{equation}\label{extr_P}
\frac{\partial P(\Omega,k)}{\partial \Omega}=0.
\end{equation}
The position of the high-frequency peak $\omega_m$ in the $S(k,\omega)$ spectra can be determined from the condition of the extremum (maximum) of function (\ref{Skw_MSC}) in frequency, which is equivalent to condition \eqref{extr_P}. Thus, the shape of the $S(k,\omega)$ spectrum for a fixed wave number $k$ will be determined by the function $P(\Omega,k)$, which is the third-degree polynomial of $\Omega$. Let us lower the order of degree of this polynomial as follows. Consider the range of variation of the frequency $\omega$ in the neighborhood of the DSF high-frequency peak corresponding to its double FWHM:
\begin{equation}
\omega_m - \Delta\omega_{1/2} \leq \omega \leq \omega_m + \Delta\omega_{1/2}.
\end{equation}
This inequality, taking into account expression \eqref{alpha}, can be rewritten in the following dimensionless form: 
\begin{equation}
\bigg|\frac{\omega-\omega_m}{\omega_m}\bigg| \leq \alpha.
\end{equation}
Next, we rewrite the last inequality through the variable $\Omega$:
\begin{equation}\label{small_param}
\bigg|\frac{\Omega-\Omega_m}{\Omega_m}\bigg| = |\overline{\Omega}|\leq \alpha(\alpha + 2),
\end{equation}
where $\Omega_m=\omega_m^2$. It follows that for the DSF spectra, which are characterized by a narrow high-frequency peak ($\alpha<<1$), the value of $\overline{\Omega}$ can be considered as a small parameter. 
If we expand the function $P(k,\Omega)$ in Taylor series over this small parameter and neglect the higher order term, we obtain
\begin{equation}\label{f1(x)} 
P(k,\Omega)\approx (3\Omega_m+A_1)\Omega^2+(A_2-3\Omega_m^2)\Omega+A_3+\Omega_m^3.
\end{equation}
Next, following the QLCA model \eqref{QLCA}, we assume that $\Omega_m\approx\Delta_1(k)+\Delta_2(k)$. Then, from expression \eqref{f(x)}, we obtain the following approximation for the DSF of the YOCP under \textit{weak decaying} collective excitations regimes:
\begin{equation}\label{Skw_SCH}
S(k,\omega) = \frac{S(k)}{\pi}\frac{B_3(k)}{\omega^4 + B_1(k)\omega^2 + B_2(k)},
\end{equation}
where
\begin{eqnarray}
B_1(k) &=&\frac{2[\Delta_2^2(k)-2\Delta_1(k)(\Delta_1(k)+5\Delta_2(k)+3\omega_0^2)]}{2\Delta_1(k)+5\Delta_2(k)+6\omega_0^2}, 
\notag \\
B_2(k) &=&\frac{2(\Delta_1(k)+\Delta_2(k))^3+3\Delta_1^2(k)(3\Delta_2(k)+2\omega_0^2)}{2\Delta_1(k)+5\Delta_2(k)+6\omega_0^2},
\notag \\
B_3(k) &=&\frac{2\sqrt{6}\Delta_{1}(k)\Delta_{2}(k)\sqrt{3\Delta_{2}(k)+2\omega_0^2}}{2\Delta_1(k)+5\Delta_2(k)+6\omega_0^2}.  \notag
\end{eqnarray}

From Eq. (\ref{Skw_SCH}), the dispersion equation for the high-frequency plasma mode of the considered system takes the form
\begin{equation}
s^{2}+\sqrt{B_1(k)+2\sqrt{B_2(k)}}s+\sqrt{B_2(k)}=0.
\end{equation}
Solving this equation gives $s(k)=\pm i\omega _{c}(k)-\delta (k)$ with dispersion for the high-frequency peak of the DSF,
\begin{equation}
\omega _{c}(k)=\frac{\sqrt{-B_1(k)+2\sqrt{B_2(k)}}}{2}  \label{omega_c}
\end{equation}
and with the dispersion for the attenuation decrement of acoustic-like collective excitations,
\begin{equation}
\delta (k)=-\frac{\sqrt{B_1(k)+2\sqrt{B_2(k)}}}{2}.
\label{sound_att}
\end{equation}
On the other hand, the position of the maximum $\omega_m(k)$ in the DSF spectra can be found from the frequency extremum condition for expression \eqref{Skw_SCH}:
\begin{equation}
\omega_m(k)=\sqrt{-\frac{B_1(k)}{2}}.
\end{equation}
Then, for the adiabatic sound velocity in the YOCP we obtain:
\begin{equation}\label{sound_speed}
c_sk = \lim_{k\rightarrow0}\omega _{c}(k) = \lim_{k\rightarrow0}\omega _{m}(k) = \lim_{k\rightarrow0}\sqrt{\Delta_1(k)}.
\end{equation}
Note that this result coincides with the expression for the adiabatic sound velocity obtained according to the Mountain scheme \cite{MokshinTMF, Mountain} when condition \eqref{gamma} is satisfied.

\begin{figure*}[tbp]
\begin{center}
\includegraphics[keepaspectratio,width=16.0cm]{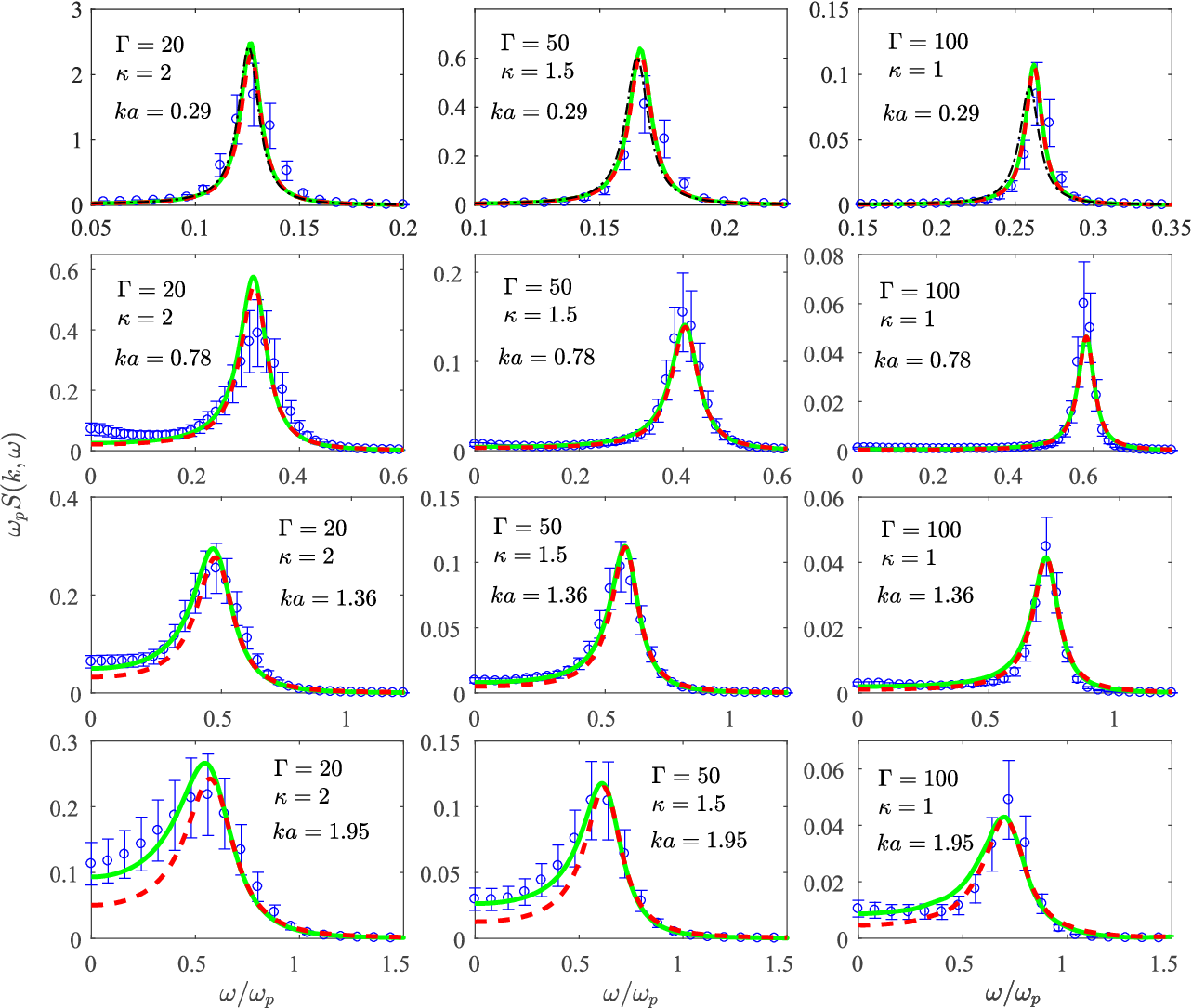}
\end{center}
\caption{Dynamic structure factor spectra of the YOCP multiplied by the plasma frequency at different values of the coupling parameter $\Gamma$, the screening parameter $\kappa$, and the reduced wave number $ka$. The theoretical results obtained from Eqs. \eqref{Skw_MSC} and \eqref{Skw_SCH}, shown in green solid and red dashed lines, respectively, are compared with the MD simulation data, shown in blue circles. The dashed lines at $ka=0.29$ are obtained within of DHO model \eqref{Skw_DHO}.}\label{Spectra}
\end{figure*}
\begin{figure*}[tbp]
\begin{center}
\includegraphics[keepaspectratio,width=16.0cm]{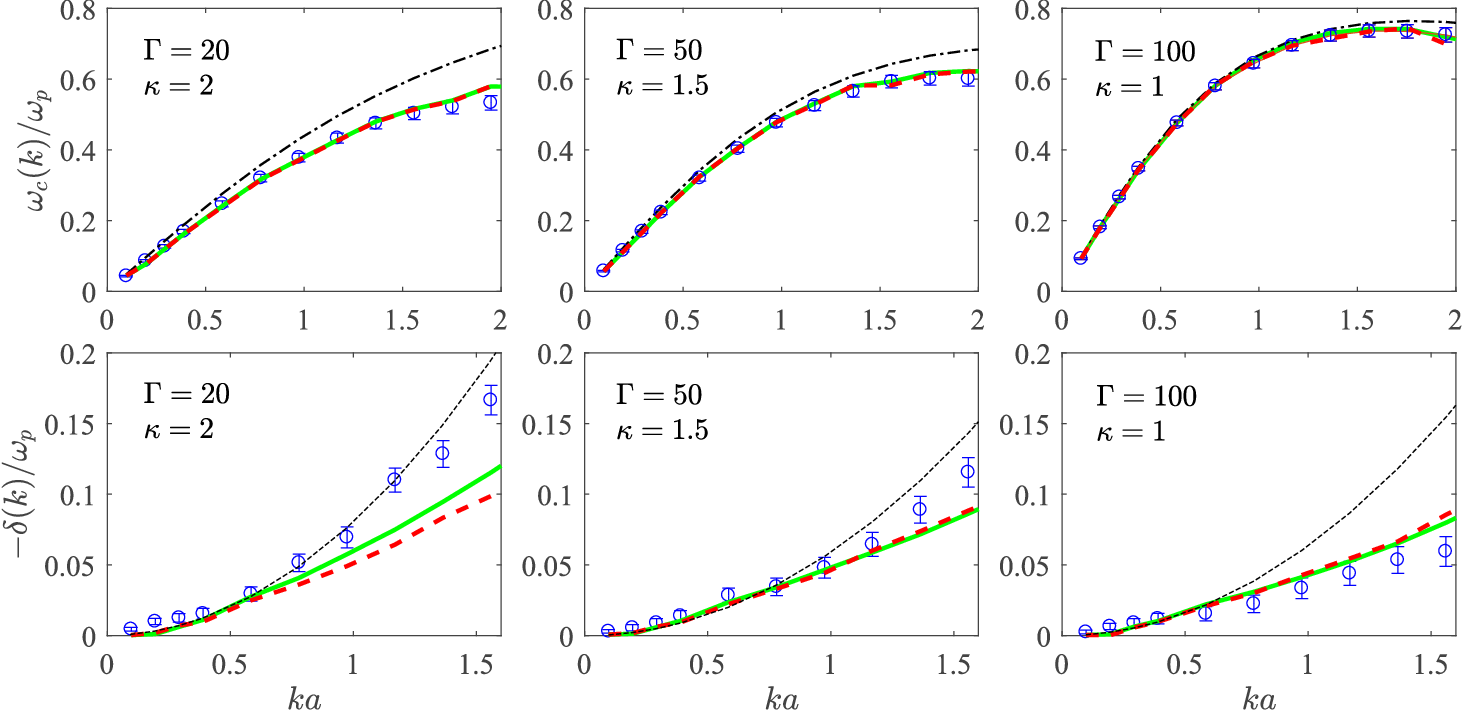}
\end{center}
\caption{Dispersions of the high-frequency peak $\omega_c(k)$ (top row) and the sound attenuation decrement $\delta(k)$ (bottom row) plotted for different values of the coupling parameter $\Gamma$ and screening parameter $\kappa$ of the YOCP. Green solid lines represent theoretical results obtained using Eqs. \eqref{omega_c} and \eqref{sound_att}, red dashed lines show calculation results using the corresponding theoretical expressions from the paper \cite{MFT_PRE}, black dashed lines are obtained within the QLCA model \eqref{QLCA}, the dashed line shows the $\sigma k^2$ dependence, and blue circles show MD simulation data.}\label{Dispersions}
\end{figure*}
The DSF in the framework of the DHO model is given by the following well-known expression \cite{Mokshin}:
\begin{equation}\label{Skw_DHO}
S^{\textrm{DHO}}(k,\omega) = \frac{S(k)}{\pi}\frac{\Delta_1(k)\Delta_2(k)\tau(k)}{[\omega^2-\Delta_1(k)]^2+[\omega\Delta_2(k)\tau(k)]^2}.
\end{equation}
Here, $\tau(k)$ is the parameter of the DHO model. Note that expression \eqref{Skw_DHO}, accounting for limiting equality \eqref{sound_speed}, becomes formula \eqref{Skw_H} for the DSF of the YOCP in the hydrodynamic limit, under the condition that the following relation is satisfied:
\begin{equation}\label{Delta_2_tau_low_k}
\lim_{k\rightarrow0}\Delta_2(k)\tau(k) = 2\sigma k^2.
\end{equation}
Since at $k\rightarrow0$ we have $\Delta_2(k)\ll\Delta_1(k)$, expression \eqref{Skw_DHO} in this limit can be written as
\begin{equation}\label{Skw_DHO_low_k}
\lim_{k\rightarrow0}S^{\textrm{DHO}}(k,\omega) = \lim_{k\rightarrow0}\frac{S(k)}{\pi}\frac{\Delta_1(k)\Delta_2(k)\tau(k)}{\omega^4-2\Delta_1(k)\omega^2+\Delta_1^2(k)}.
\end{equation}
On the other hand, in this limit, the coefficients $B_1(k)$, $B_2(k)$ and $B_3(k)$ of expression \eqref{Skw_SCH} take the form
\begin{equation}\label{B1_low_k}
\lim_{k\rightarrow0}B_1(k)=-2\Delta_1(k),
\end{equation}
\begin{equation}\label{B2_low_k}
\lim_{k\rightarrow0}B_2(k)=\Delta_1^2(k),
\end{equation}
\begin{equation}\label{B3_low_k}
\lim_{k\rightarrow0}B_3(k)=\frac{2\sqrt{3}}{3\omega_0}\Delta_1(k)\Delta_2(k).
\end{equation}
Comparing Eqs. \eqref{Skw_SCH} and \eqref{Skw_DHO_low_k}, taking into account limiting relations \eqref{B1_low_k}-\eqref{B3_low_k}, we obtain the following simple formula for the parameter $\tau(k)$ in the limit $k\rightarrow0$:
\begin{equation}\label{tau}
\tau = \frac{2\sqrt{3}}{3\omega_0}.
\end{equation}
Further, since the sound attenuation coefficient $\sigma$ in the DHO model is related to the parameter $\tau$ and to the second-order frequency relaxation parameter $\Delta_2(k)$ through relation \eqref{Delta_2_tau_low_k}, we obtain
\begin{equation}\label{damp_coeff}
\sigma  = \frac{\sqrt{3}}{3\omega_0}\lim_{k\rightarrow0}\frac{\Delta_2(k)}{k^2}.
\end{equation}
This relation expresses the fact that such a dynamic characteristic of the YOCP as the sound attenuation coefficient is directly determined in the proposed theoretical model through the structural characteristic of the system: the radial distribution function $g(r)$.

\section{Results and discussion}
Now, we compare the theoretical results with the MD simulation data. Figure \ref{Spectra} shows the DSF spectra of the YOCP that were obtained using Eqs. \eqref{Skw_MSC} and \eqref{Skw_SCH}, as well as the MD simulation results. As can be seen, the DSF  spectra of the YOCP constructed using Eq. \eqref{Skw_SCH} do not differ much from the spectra obtained from analytical expression \eqref{Skw_MSC}. From this we can conclude that the propagating \textit{weak decaying} collective excitations that are characteristic of the YOCP in the range of wave numbers $0<k\lesssim k_m/2$ can be described within the DHO-like model \eqref{Skw_SCH}. At the same time, this model requires knowledge of only the parameters $\Gamma$ and $\kappa$, as well as information about the structure. The maximum discrepancy is observed at $\Gamma=20$ and $\kappa=2$ for collective excitations with the reduced wave number $ka = 1.95$, which are characterized by the largest decay. Note that the upper limit of the considered values of the reduced wave number $ka = 1.95$ corresponds to the spatial scale $L\approx2l$, where $l$ is the average interparticle distance. Thus, the presented theoretical model describes the DSF of the YOCP on a wide spatial scale: from extended hydrodynamics to interparticle distances. The DSF spectra constructed within DHO model \eqref{Skw_DHO} using expression \eqref{tau} for the parameter $\tau(k)$ practically coincide with the spectra constructed within of the theoretical model for DSF of the YOCP developed in the present work (upper row of Fig. \ref{Spectra}).

\begin{table}[h]\caption{The reduced sound attenuation coefficient in the YOCP $\overline{\sigma}=\sigma/(\omega_pa^2)$, calculated using Eq. \eqref{damp_coeff} (the “Th”\ index) and obtained from MD simulation data (the “MD”\ index).}
\begin{center}
\begin{tabular}{c c c c}
\hline
$\Gamma$ & $\kappa$ & $\overline{\sigma}_{\textrm{Th}}$ & $\overline{\sigma}_{\textrm{MD}}$ \\
\hline
20 \,&\, 1.0 \,&\, 0.062 $\pm$ 0.009\,& \,0.060 $\pm$ 0.011\\
50 \,&\, 1.0 \,&\, 0.061 $\pm$ 0.007 \,& \,0.058 $\pm$ 0.009\\
100 \,&\, 1.0 \,&\, 0.065 $\pm$ 0.003 \,& \,0.062 $\pm$ 0.006\\
20 \,&\, 2.0 \,&\, 0.065 $\pm$ 0.007 \,& \,0.067 $\pm$ 0.009\\
50 \,&\, 2.0 \,&\, 0.048 $\pm$ 0.003 \,& \,0.046 $\pm$ 0.005\\
100 \,&\, 2.0 \,&\, 0.052 $\pm$ 0.008 \,& \,0.055 $\pm$ 0.009\\
\hline
\end{tabular}
\end{center}
\end{table}\label{Tab.1}

Figure \ref{Dispersions} shows the dispersions of the DSF high-frequency peak $\omega_c(k)$ and the attenuation decrement dispersions $\delta(k)$ obtained using Eqs. \eqref{omega_c} and \eqref{sound_att}, respectively. Here we also present the plots that are obtained using the corresponding expressions from Ref. \cite{MFT_PRE} and the results of MD simulations. As in the case with the DSF spectra, the theoretical curve for the dispersion of the attenuation decrement $\delta(k)$ diverges maximally from the simulation data for the YOCP state with $\Gamma=20$ and $\kappa=2$. At the same time, the results of calculations of the DSF high-frequency peak  dispersions using Eq. \eqref{omega_c} are in excellent agreement with the MD simulations data for all considered thermodynamic states of the YOCP. The sound attenuation coefficient calculations results for the YOCP using Eq. \eqref{damp_coeff} at different ($\Gamma, \kappa$) states are shown in Table I. The calculation results from MD simulation data are also given here. It can be seen that the results of theory and simulations are in good agreement. Note that since approximation \eqref{gamma} is valid for the YOCP at the considered ($\Gamma, \kappa$) states, the sound attenuation coefficient turns out to be proportional to the shear viscosity coefficient \cite{Mithen_PRE_2011}. According to several works \cite{DonkoPRE2008, MithenContrib2012, Saigo, KhrapakAIP2018}, the $\Gamma$ dependence of the shear viscosity coefficient of the YOCP for $\kappa=2$ has a minimum at $\Gamma\approx50$, which is consistent with the data from Table I.

\section{Conclusion}
In conclusion, we note that the model of \textit{weak decaying} collective excitations in the Yukawa one-component plasmas presented in this work can serve as a simple generalization of the known quasilocalized charge approximation. The developed model is based on the self-consistent relaxation theory of collective dynamics and the previously established correlation relations linking the parameters characterizing three- and four-particle correlations with the parameter characterizing pair correlations in the Yukawa plasmas. Calculations of the spectra of the dynamic structure factor and dispersion characteristics using the obtained analytical expressions give results consistent with the data of molecular dynamics simulation for all considered ($\Gamma, \kappa$) states. It is shown that the presented theoretical model for the dynamic structure factor in the limit of small wave numbers corresponds to the damped harmonic oscillator model. This correspondence allows one to obtain the simple analytical expression for the sound attenuation coefficient of the Yukawa one-component plasmas. The results of calculations using this expression agree with the literature data for the shear viscosity of the Yukawa one-component plasmas.

\section{ACKNOWLEDGEMENTS} The work was carried out on the basis of the grant provided by the Tatarstan Academy of Sciences in 2024 for the implementation of fundamental and applied research work in scientific and educational organizations, enterprises and organizations of the real sector of the economy of the Republic of Tatarstan. 
%The authors are grateful to I. M. Tkachenko and S. A. Khrapak for helpful discussions.

\section{APPENDIX: MOLECULAR DYNAMICS SIMULATION DETAILS} MD simulations of the YOCP were performed in the LAMMPS package \cite{LAMMPS} for the equilibrium configuration of the YOCP at $\Gamma = 20, 50, 100$ and $\kappa = 1, 1.5, 2$ in the NVT ensemble. The simulation cell contained 64 000 particles interacting through the Yukawa potential. Periodic boundary conditions in all directions were applied to the cell. The equations of motion of the particles were integrated using the velocity-based Verlet algorithm with a time integration step $\tau=0.01/\omega_p$.

\end{document}